\documentstyle[prl,aps,epsf]{revtex}
\begin{document}
\draft
\twocolumn[
\hsize\textwidth\columnwidth\hsize\csname@twocolumnfalse\endcsname
\preprint{}
\title {Theory of Magnetic Field Induced Spin Density Wave
in High Temperature Superconductors}
\author{Yan Chen and C. S. Ting}
\address{Texas Center for Superconductivity and Department of
Physics, University of Houston, Houston, TX 77204}
\maketitle
\begin{abstract}
{The induction of  spin density wave (SDW) and charge density wave (CDW)
orderings
in the mixed state of high $T_c$ superconductors (HTS) is investigated by
using the self-consistent Bogoliubov-de Gennes equations based upon an
effective model Hamiltonian with competing SDW and $d$-wave
superconductivity interactions.  For optimized doping sample, the modulation of the induced SDW and 
its associated CDW is determined by the vortex lattice and their patterns obey the four-fold symmetry. 
By deceasing doping level, both SDW and CDW show 
quasi-one dimensional like behavior, and the CDW has  a period just
half that of the SDW along one direction.  From the calculation of the
local density of states (LDOS),  we found that the majority of the
quasi-particles inside the vortex core are 
localized. All these results are consistent with several recent
experiments on HTS}.
\end{abstract}
]

\narrowtext
Intensive efforts have been focused recently on the nature of vortex
core excitations and the possible induction of SDW
and other phases in the mixed state of high $T_c$ superconductors (HTS).
Experiments from the scanning tunneling
microscope(STM)~\cite{Maggio,Pan00,Hoff1}, neutron
scattering~\cite{Katano,Lake01,Lake02} and nuclear magneticresonance(NMR)~\cite{NMR1} provided 
vital informations on these topics.
For example, according to the neutron scattering experiment by Lake {\em et
al.}~\cite{Lake01},  a remarkable antiferromagnetism or SDW appears
in the optimally doped La$_{2-x}$Sr$_x$CuO$_4$ when a strong magnetic
field  is applied.  Very recently, STM measurement by Hoffman{\em et al.}~\cite{Hoff1}
studied the LDOS in the mixed states of
optimally doped
Bi$_2$Sr$_2$CaCu$_2$O$_{8+ \delta}$, and they found that associated with SDW, 
an inhomogeneous CDW exist both inside and outside the vortex cores.
The coexistence of d-wave superconductivity (DSC)
and SDW orders was theoretically
studied by several groups in the absence of a magnetic field
~\cite{Emery,Bale98,Kyung,Mart00}. In the presence of a magnetic field, this problem was studied 
by the SO(5) theory~\cite{Arovas97} and also by a Ginzburg-Landau approach~\cite{Demler}. 
The LDOS at the vortex core in a pure DSC were first calculated by authors in Ref.(14).
With induced antiferromagnetic (AF) order, the LDOS was investigated without the contribution of the quasiparticles
~\cite{Arovas97,Demler}, and by a mean field study~\cite{Zhu01}. 

Although the competition between the SDW and DSC in a magnetic field was
previously examined, the nature of the induced SDW and  its
spatial variation have not been addressed in such  detail as to
compare with the experiments. In this paper, we shall adopt
the method described in previous papers~\cite{Wang95,Zhu01}
to examine the possible induction of extended SDW and CDW orders in the mixed state of HTS, and
their nature in under-, optimally and overdoped samples.  In order to
simplify the numerical calculation, we shall assume  a square vortex
lattice for  the mixed state and a strong magnetic field B such that
$ \lambda \gg b \gg \xi$ with $\lambda$ as the London  penetration depth,
$\xi$ the coherence length and $b$  the vortex lattice constant.
Under this condition, the applied magnetic field B can be regarded as a
constant throughout the sample.  Our calculation is based upon an
effective model Hamiltonian with competing SDW and DSC orderings. In the
absence of a magnetic field B and for the optimally (hole)  doped sample
$x=0.15$, the parameters are chosen in such a way that the SDW
ordering is completely suppressed and only the DSC ordering prevails. When
the system is in the mixed state driven by a magnetic field B, an
inhomogeneous SDW is
induced. We found that the structure of the induced SDW is determined  and
pinned by the underlying vortex lattice. 
 For optimized doping sample, the modulation of the induced SDW and 
its associated CDW is determined by the vortex lattice and their patterns obey the four-fold symmetry. 
With  the same set parameters and  B=0,  SDW, DSC and CDW stripes appear in the
underdoped sample $(x=0.12)$.  In the presence of a strong B, all 
order parameters pattern show remarkable anisotropic behavior along the $x$ and $y$ directions, and the 
CDW has a period just half that of the SDW along one direction. Increasing the SDW 
interaction strength can also lead to quasi-one dimensional pattern.
The LDOS near the vortex core has also been
calculated. It shows an asymmetric double peaks around $E=0$ in agreement
with the results of Ref.(15) and experiments~\cite{Maggio,Pan00}.  In addition our results also
show a small gap at $E=0$ indicating the presence of the SDW order in the bulk
sample.  From the spatial profile of the LDOS,  we conclude that almost all
the quasiparticles inside the vortex core are localized.

Let us begin with a phenomenological model in which interactions
describing  both DSC and SDW order parameters in a two-dimensional lattice are
considered. The effective Hamiltonian can be written as:
\begin{eqnarray}
H&=&\sum_{{\bf i,j},\sigma} - {t_{\bf i,j}} c_{{\bf
i}\sigma}^{\dagger}c_{{\bf j}\sigma}
+\sum_{{\bf i},\sigma}( U n_{{\bf i} {\bar {\sigma}}} -\mu)c_{{\bf i}\sigma}^{\dagger} c_{{\bf i}\sigma}
\nonumber \\
&&+\sum_{\bf i,j} ( {\Delta_{\bf i,j}} c_{{\bf i}\uparrow}^{\dagger}
c_{{\bf j}\downarrow}^{\dagger} + h.c.)\;.
\end{eqnarray}
The summation here is over the nearest neighbors sites.
$c_{{\bf i}\sigma}^{\dagger}$ is the electron creation operator and $\mu$
is the chemical potential.
In the presence of magnetic field B, the hopping integral can be expressed
as  $ t_{\bf i,j}= t_0 e^{i  \frac{ \pi}{\Phi_{0}}
\int_{{\bf r}_{\bf j}}^{{\bf r}_{\bf i}} {\bf A}({\bf r})\cdot d{\bf r}}$
for the nearest neighboring sites $(i,j)$. For simplicity, we have set the
next-nearest neighbor hopping
equal to zero. $\Phi_0=h/2e$ is the superconducting flux quanta.
Here we choose Landau gauge
${\bf A}=(-By,0,0)$ with $y$ as the $y$-component of the position vector
{\bf r}.    The two possible orders in cuprates are
SDW and DSC which have the following definitions respectively:
$\Delta^{SDW}_{\bf i} = U \langle c_{{\bf i} \uparrow}^{\dagger}
c_{{\bf i} \uparrow} -c_{{\bf i} \downarrow}^{\dagger}c_{{\bf i}
\downarrow} \rangle$
and   $\Delta_{\bf i,j}=V_{DSC} \langle
c_{{\bf i}\uparrow}c_{{\bf j}\downarrow}-c_{{\bf i}
\downarrow} c_{{\bf j}\uparrow} \rangle /2$.
In the above expressions, $U$ and $V_{DSC}$ are respectively
the interaction strengths for SDW and DSC orders.
The mean-field Hamiltonian (1) can be diagonalized by solving
the resulting Bogoliubov-de Gennes equations self-consistently
\begin{equation}
\sum_{\bf j} \left(\begin{array}{cc}
{\cal H}_{\bf i,j}& \Delta_{\bf i,j} \\
\Delta_{\bf i,j}^{*} & -{\cal H}_{\bf i,j}^{*}
\end{array}
\right)
\left(\begin{array}{c} u_{\bf j}^{n} \\
v_{\bf j}^{n}
\end{array}
\right)
=E_{n}
\left(
\begin{array}{c}
u_{\bf i}^{n} \\
v_{\bf i}^{n}
\end{array}
 \right)\;,
\end{equation}
where the single particle Hamiltonian ${\cal H}_{\bf i,j}^{\sigma}=
-t_{\bf i,j} +(U n_{{\bf i} \bar{\sigma}} -\mu)\delta_{\bf ij}$, and\begin{equation}
n_{{\bf i} \uparrow} = \sum_{n} |u_{\bf i}^{n}|^2 f(E_{n}),
\end{equation}
\begin{equation}
n_{{\bf i} \downarrow} = \sum_{n} |v_{\bf i}^{n}|^2 ( 1- f(E_{n})),
\end{equation}
\begin{equation}
\Delta_{\bf i,j} = \frac{V_{DSC}} {4} \sum_{n}
(u_{\bf i}^{n} v_{\bf j}^{n*} +v_{\bf i}^{*} u_{\bf j}^{n}) \tanh
\left( \frac{E_{n}} {2k_{B}T} \right) ,\end{equation}
with $f(E)$ as the Fermi distribution function and the electron density
$n_{\bf i}= n_{{\bf i} \uparrow} + n_{{\bf i} \downarrow}$. The DSC order
parameter at each site $i$ is $\Delta^{D}_{\bf i}=
(\Delta^{D}_{\bf i+e_x,i} + \Delta^{D}_{\bf i-e_x,i} - \Delta^{D}_{\bf
i,i+e_y}
-\Delta^{D}_{\bf i,i-e_y})/4$ where
$ \Delta^{D}_{\bf i,j} = \Delta_{\bf i,j} exp[ i {
\frac{\pi}{\Phi_{0}}
\int_{{\bf r}_{\bf i}}^{({\bf r}_{\bf i}+{\bf r}_{\bf j})/2 } {\bf A}({\bf
r}) \cdot d{\bf r}}]$ and ${\bf e}_{x,y}$ denotes the unit vector along $(x,y)$ direction. 
The main procedure of self-consistent calculation
is given below:  For a given initial set of parameters $n_{{\bf i} \sigma}$ and
$\Delta_{\bf i, j}$, the Hamiltonian is numerically diagonalized
and the  electron wave functions obtained are used
to calculate the new parameters for the next iteration step.
The calculation is repeated until the relative difference of order parameter
 between two consecutive iteration step is less than $10^{-4}$.
The solutions corresponding to various doping concentrations can be obtained 
by varying the chemical  potential.

In the following calculation, the length and energy are measured in units
of the lattice constant $a$ and the hopping integral $t_0$ respectively.
We need to point out that the induction of internal magnetic field by the  supercurrent 
around the vortex core is so small comparing with the external magnetic field that we can 
safely adopt the unfiorm magnetic field distribution approximation.
We follow the standard procedures~\cite{Wang95,Zhu01} to
introduce magnetic unit cells,
where each unit cell accommodates two superconducting flux quanta.
By introducing the quasi-momentum of the magnetic Bloch state, we obtain
the wave function under the periodic boundary condition whose region
covers many unit cells.
The related parameters are chosen as the following: The DSC coupling
strength is $V_{DSC}=1.2$, the linear dimension of the unit
cell of the vortex lattice is chosen as $N_x \times N_y = 40 \times 20$
sites  and the number of the unit cells $M_x \times M_y = 20 \times 40$.
This choice corresponds the magnetic field $B \simeq 37 T$.

At optimal doping and $B=0$, the important
qualitative characteristic of the quasiparticle states is identical to that of a pure
d-wave superconductor. At the first step, the magnitudes of
parameters are selected to fulfill such requirement that
the AF order is completely suppressed and only DSC
prevails in the absence of a magnetic field. Here we choose the hole
doping $x= 0.15$ and $U=2.4$.
Our calculation is performed at very low temperature where the vortex
structure is almost independent of temperature.
In Fig. 1(a) we plot the typical configuration of vortex structure.
The DSC order parameter vanishes at the
vortex core center and recovers its bulk value at a couple of coherence
lengths away from the center.
By comparing it with the vortex structure of a pure DSC, the size of the
vortex core here is noticed  to be enlarged.
The centers of the two vortex cores are at sites $(10,10)$ and $(30,10)$. 
Fig. 1(b) displays the spatial distribution of the
staggered magnetization of the induced SDW order as defined by
$M_{\bf i}^{s}=(-1)^{i} \Delta^{SDW}_{\bf i}/U$. It is obvious that
the AF order exists both inside and outside the vortex cores,
and behaves like a two-dimensional SDW with the same wave length
in the $x$ and $y$ directions.
The induced  SDW order reaches its maximum value at the vortex core center
and the magnitude of  its spatial variation still holds the fourfold symmetry as
that of the pure DSC case. The order of DSC and SDW coexist throughout the whole sample.
The appearance of the SDW order around the vortex cores strongly affects
the spatial profile of the local
electron density distribution, which can be represented by a weak CDW as shown in Fig. 1(c).
The remarkable enhancement of electron density (or depletion of the hole
density)  is presented at the vortex core center. 
It is easy to observe that the four-fold symmetry holds for both SDW and 
CDW with the same wavelength $20 a$.

To have a deeper understanding of the above results, the case 
of an underdoped sample $(x=0.12)$ is examined.  After the calculation is performed at $B=0$, 
we found that  DSC, SDW and
CDW orderings have the stripe structuress, which is consistent with
stripe phase results reported by Martin {\em et al.}~\cite{Mart00}.
We have compared the free energy between stripe phase solution
with a uniform AF solution and found that the free energy of the stripe
phase is always lower. 
In the presence of a strong magnetic field B, the
profile of DSC order parameter is shown in Fig. 2(a). The radius of vortex is further enlarged than in Fig. 1(a).
The qualitative features of Fig. 2 are quite different from those in Fig.1 and 
the spatial variation  of the SDW becomes quasi-one dimensional. Its periods of oscillations
are fixed by the vortex lattice similar to the case of $x=0.15$. We also noticed
that the AF order is much enhanced in the mixed state as compared with its
values in the stripe phase at $B=0$, in agreement with Katano's
experiment~\cite{Katano}. The results are presented in Fig. 2(b) and Fig. 2(c), where
the anisotropy in the magnitudes of SDW and CDW along the $x$ and $y$ directions shows
quasi-one-dimensional behavior. The periods  of SDW and CDW are respectively $20a$ and $10a$. 
These results are in qualitative agreement with the observations of  Lake {\em et al.}~\cite{Lake02}
where a magnetic field induced striped AF order was observed in underdoped
sample. We also have calculated the case for overdoped sample ($x =0.20$) with
the same set of parameters. For $B=0$, the SDW order is completely
suppressed and the DSC order is homogeneous in real space. When B is strong, the SDW order
does not show up even inside the vortex core.

We notice that the periods of SDW and CDW obtained from the experiments~\cite{Hoff1,Lake01}
are respectively $8a$ and $4a$ for optimal doping sample. With the present band parameters, we
are not able to obtain these numbers. However, 
the experimental values could be obtained by tuning $U$, doping or including a 
next-nearest neighbor hopping term.
For larger $U$ case, the configurations of SDW and CDW exhibit stripe-like behavior 
along the $x-$(or $y-$) direction. The periodicity of CDW is always half that of SDW. 
For example when $U=3$ and $x=0.20$, 
the periodicity of SDW and CDW are respectively $10a$ and $5a$. 
But under this condition, the AF stripe phase would appear in the optimally doped sample.

Next  we present  the calculation of  the LDOS near the center of
vortex  core for hole doping $x=0.15$. The LDOS is given by:
\begin{equation}
\rho_{\bf i}(E) = -\sum_{n}[\vert
u_{\bf i}^{n}\vert^{2}
f^{\prime}(E_{n}-E) +\vert v_{\bf
i}^{n}\vert^{2}f^{\prime}(E_{n}+E)]\;,
\end{equation}
It also measures the differential tunnel conductance, which could be
measured by STM experiments.
In this case, the thermally broadening effect has not been considered here and the 
temperature is fixed at $T=0.01$.
We plot the LDOS at the core center vs the quasi-particle energy measured
from the Fermi level.
For comparison, we have also displayed the LDOS at the site
between two next-nearest neighboring vortex cores.  The results are shown in Fig. 3.
At the core center, the coherent peaks due to the gap edges of the
superconductor at $B=0$ are suppressed, and two asymmetric peaks of the
vortex states appear slightly above and below $E=0$. The spectrum agrees
qualitatively with the experiment for YBCO~\cite{Maggio}, and that of Ref.(15).
But a  closer inspection reveals that the distance between these two
peaks becomes slightly larger  than that in Ref.(15). 
The peak at about $E=0.8$ seems to be the
characteristic of the bulk SDW order and the band structure effect. 
 It is  easy to  notice that the magnitude of the  
LDOS at $E=0$ approaches zero which indicates a bulk SDW gap.
From the spatial distribution of LDOS at $E=0$ (see Fig. 4(a)), we find that LDOS 
is enhanced along the $x=y$ and $x=-y$ or the diagonal directions from the center of the vortex core.
 This 'star'-like behavior indicates the small number of quasiparticles very close to $E=0$ are 
extended~\cite{zhumap}. We also present the result (see Fig. 4(b)) for the peak on the left of $E=0$ in Fig. 3. 
Its profile displays an obvious localized shape even though it decays somewhat slower along the $x$ and $y$ 
directions than the diagonal directions. The peak on the right at  $E \simeq 0.16$ has a similar spatial distribution 
which is not shown here. From Fig. 4(b), we conclude that the majority of quasiparticles inside the vortex core 
with energies near or at the two asymmetric peaks are in fact localized, 
in agreement with experiments~\cite{Hoff1,Lake01}.

In conclusion, we have studied the induction of SDW and CDW
orders in optimally and  underdoped  HTS  by a strong magnetic field.
Consider only the nearest neighbor hopping term, 
the spatial variations of DSC, SDW and CDW orders have been numerically
presented in Fig. 1 and Fig. 2 and stripe-like structure exhibits for underdoped sample.  
The  LDOS  near the vortex center have also been calculated. We show that almost all the
quasi-particles inside the core are localized, that is very different from
case for a pure DSC vortex~\cite{zhumap}. These results are consistent with
recent STM, neutron scattering and NMR experiments on HTS.
Finally we would like to emphasize that although our self-consistent BdG
calculation based upon a mean-field approach  tends to overestimate the
stability of the SDW phase, the qualitative features of our results should
still be valid, particularly in view of their favorable comparisons with
experiments.

We are grateful to Dr. J.-X. Zhu and Prof.  S. H. Pan for useful
discussion and Prof. J. C. Davis for showing us the results of Ref.(3) before its publication. 
This work was
supported by a grant from  the Robert A. Welch Foundation and by the Texas
Center for Superconductivity at the University of Houston through the
State of Texas, and by a Texas ARP grant(003652-0241-1999).

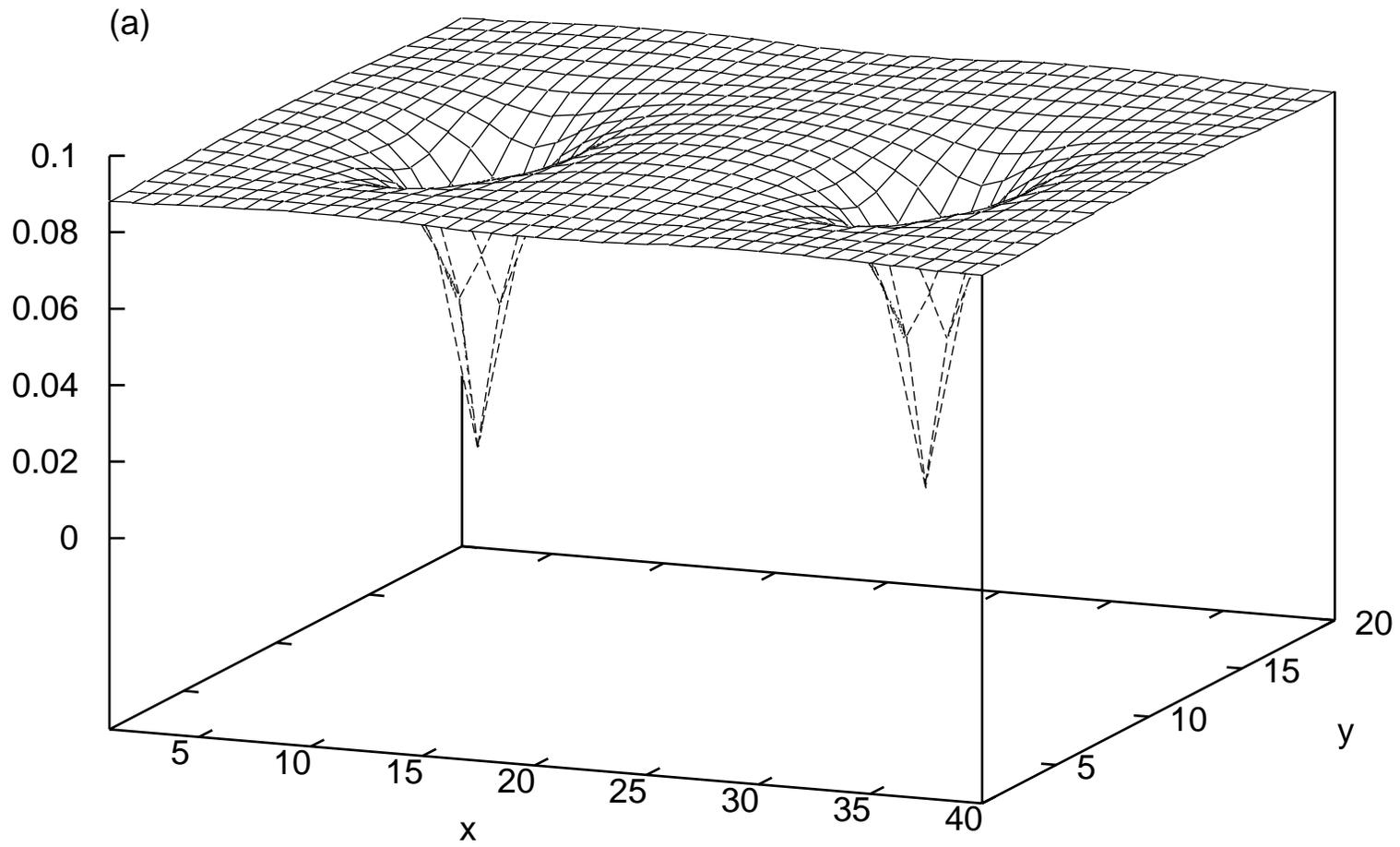
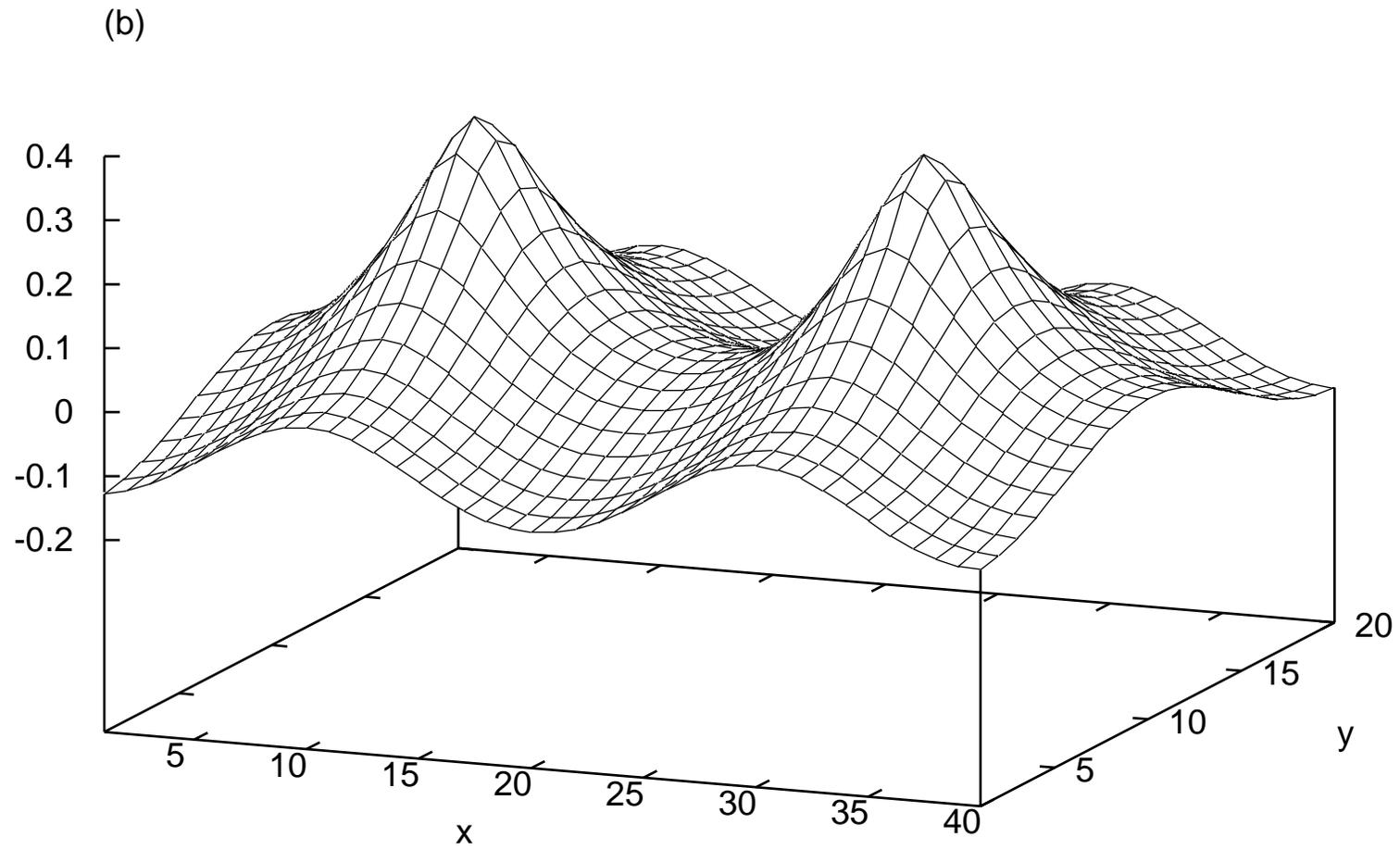
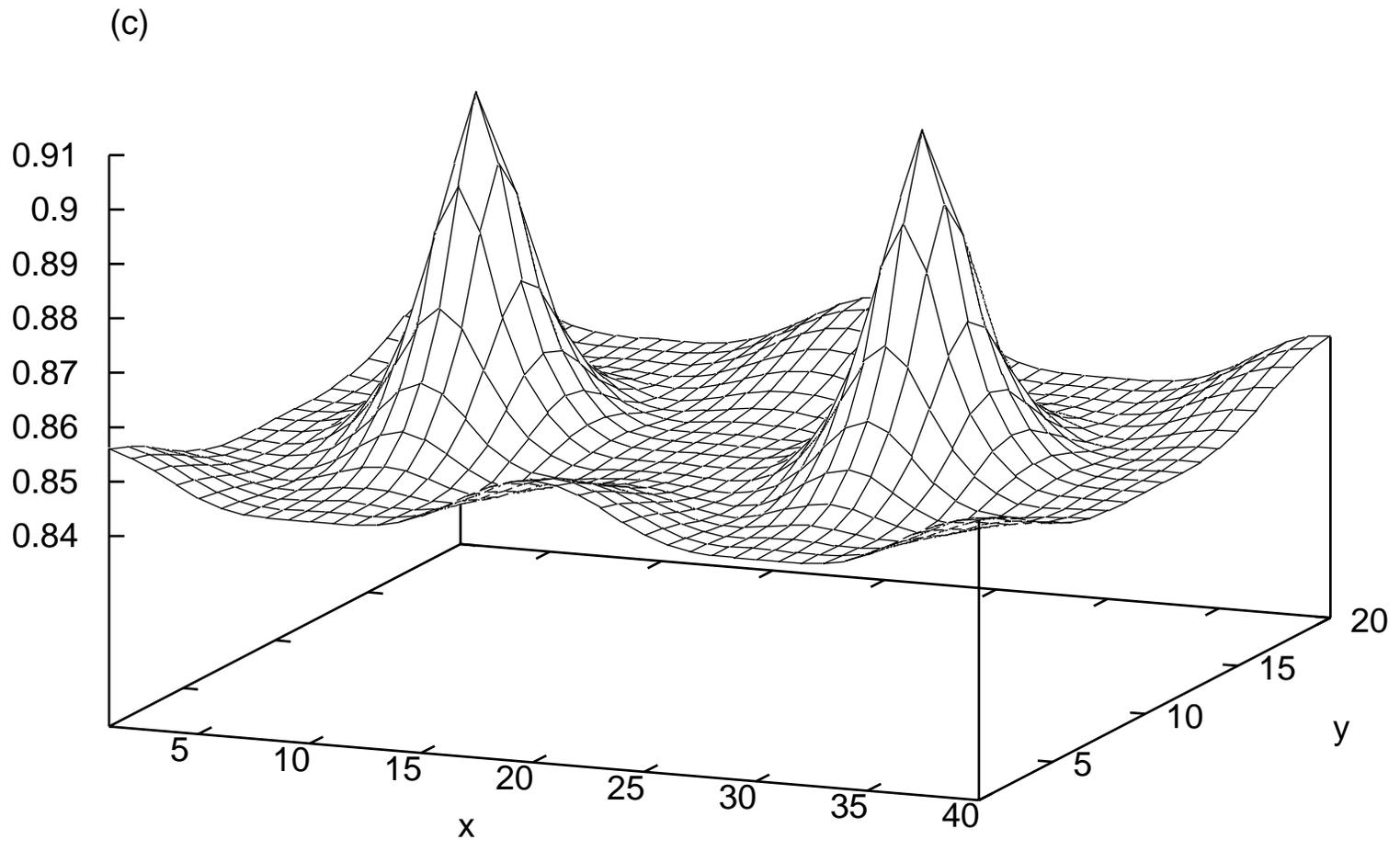
\begin{figure}\caption[*] {The amplitude distribution of the DSC order parameter $\Delta_{\bf i}^{D}$ (a), 
the staggered magnetization $M_{\bf i}^{s}$ (b), and the electron density $n_{\bf i}$ (c) in one magnetic unit cell. 
The size of the cell is $40 \times 20$, corresponding to a magnetic field $H= {\Phi_0}/(20 \times 20)$. The strength of the on-site repulsion $U=2.4$ and the averaged electron 
density $ \bar {n}=0.85$.}\label{1}
\end{figure}

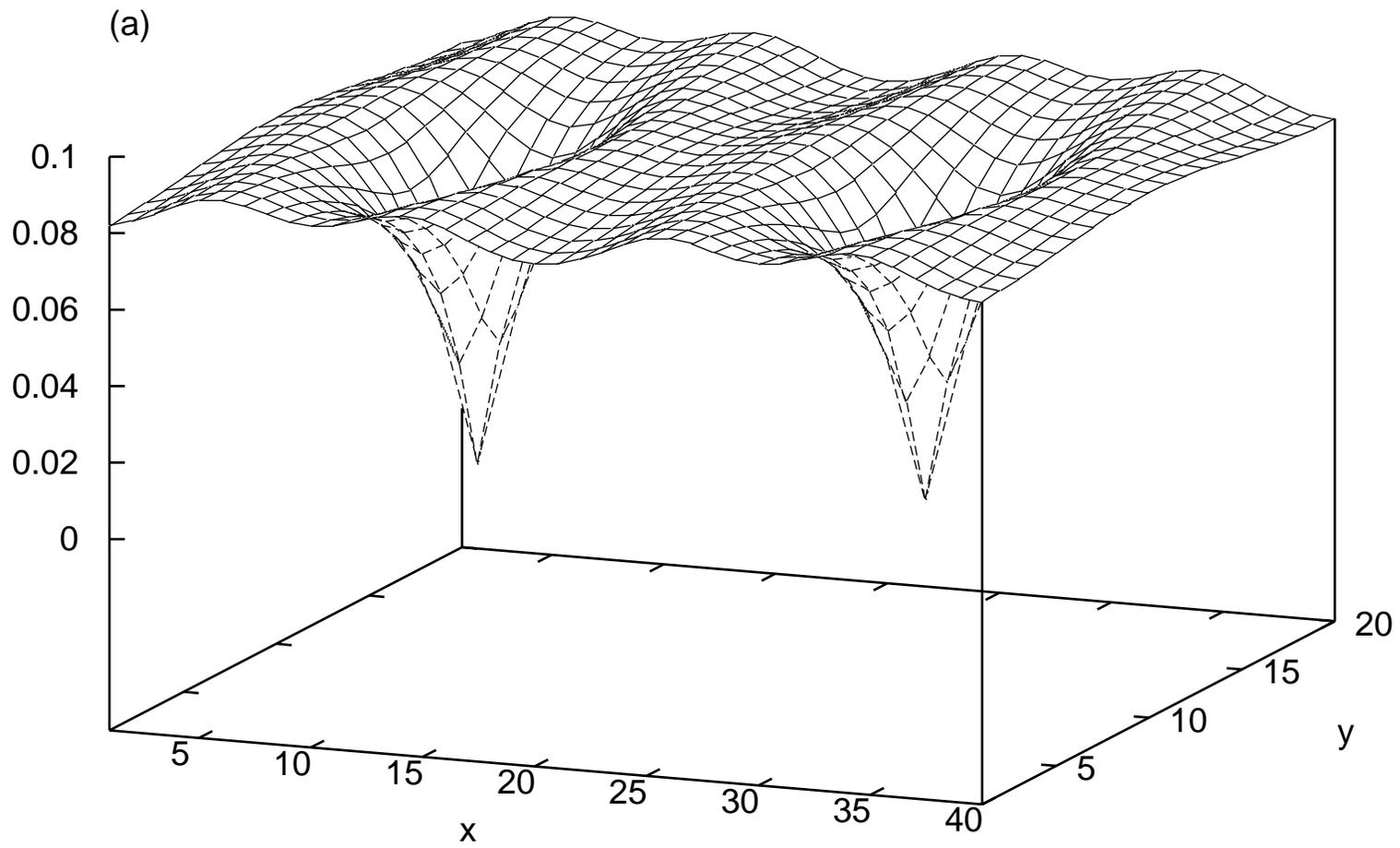
\begin{figure}\caption[*] {The amplitude of the DSC order parameter $\Delta_{\bf i}^{D}$ 
(a), the staggered magnetization $M_{\bf i}$ (b), and the electron density $n_{\bf i}$ (c) 
in one magnetic unit cell. The averaged electron density 
$\bar {n} =0.88$. The other parameter values are the same as Fig. 1.}\label{2}
\end{figure}

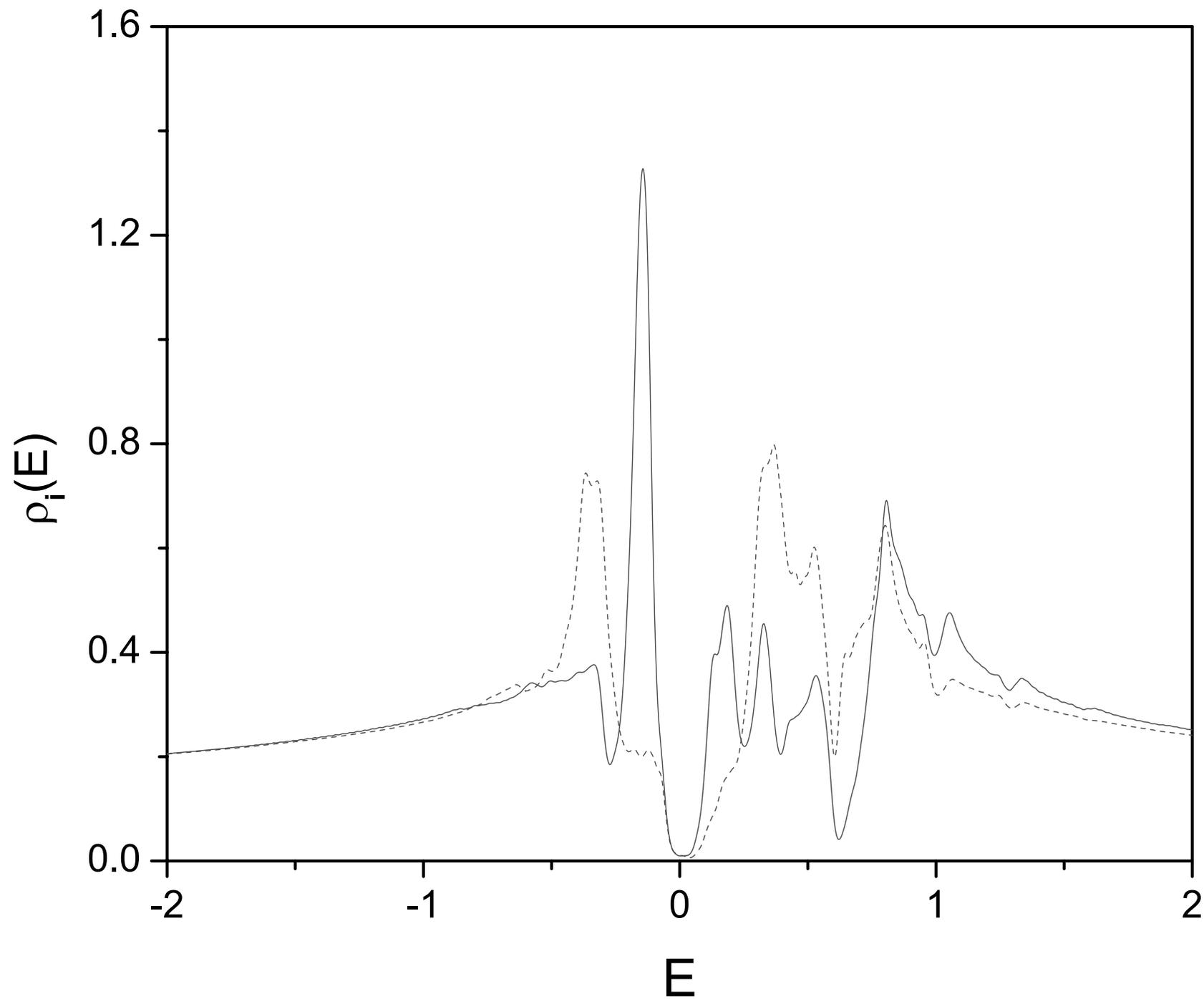
\begin{figure}\caption[*] { Quasiparticle LDOS profiles at the center of the vortex core are shown by solid lines. 
The dashed lines show the LDOS at the site midway between two next-nearest neighbor vortices.
The averaged electron density $\bar {n} =0.85$.}\label{3}
\end{figure}

\begin{figure}\caption[*] { The spatial distribution of LDOS in a vortex unit cell at the energy $E =0$ 
a) and $E =-0.126 $ (b).  }\label{4}
\end{figure}


\begin{thebibliography}{18}
\bibitem{Maggio} I. Maggio-Aprile {\em et al.}, Phys. Rev. Lett.
{\bf 75}, 2754 (1995).
\bibitem{Pan00} S. H. Pan {\em et al.}, Phys. Rev. Lett. {\bf 85}, 1536
(2000).
\bibitem{Hoff1} J. E. Hoffman {\em et al.},  Science {\bf 295}, 466 (2002).
\bibitem{Katano} S. Katano et al., Phys. Rev. B. {\bf 62}, R14677 (2000).
\bibitem{Lake01} B. Lake {\em et al.}, Science {\bf 291}, 1759 (2001);
 B. Lake {\em et al.},  cond-mat/0104026.
\bibitem{Lake02} B. Lake {\em et al.}, Nature {\bf 415}, 299 (2002).
\bibitem{NMR1} V. F. Mitrovic {\em et al.}, Nature {\bf 413}, 501 (2001).
\bibitem{Emery} V. J. Emery and S. A. Kivelson, Nature {\bf 374}, 434 (1995);V. J. Emery, S. A. Kivelson, 
and J. M. Tranquada, Proc. Natl. Acad. Sci. USA {\bf 96}, 8814 (1999);
\bibitem{Bale98} L. Balents, M. P. A. Fisher, and C. Nayak, Int. J. Mod.
Phys. B {\bf 12}, 1033 (1998).
\bibitem{Kyung} Bumsoo Kyung, Phys. Rev. B. {\bf 62}, 9083 (2000).
\bibitem{Mart00} I. Martin, {\em et al.}, Int. J. Mod. Phys. B {\bf 14}, 3567 (2000).
\bibitem{Arovas97}  D. P. Arovas {\em et al.}, Phys. Rev. Lett. {\bf 79}, 2871 (1997);  
J. -P. Hu, and S. C. Zhang, cond-mat/0108273.
\bibitem{Demler} E. Demler, S. Sachdev, and Y. Zhang, Phys. Rev. Lett. {\bf 87},  067202 (2001); 
A. Polkovnikov, {\em et al.}, cond-mat/0110329.
\bibitem{Wang95} Y. Wang and A. H. MacDonald, Phys. Rev. B {\bf 52}, R3876 (1995).
\bibitem{Zhu01} Jian-Xin Zhu, and C. S. Ting, Phys. Rev. Lett. {\bf 87}, 147002 (2001).
\bibitem{zhumap} Jian-Xin Zhu, C. S. Ting, and A. V. Balatsky, cond-mat/0109503.

\end{thebibliography}
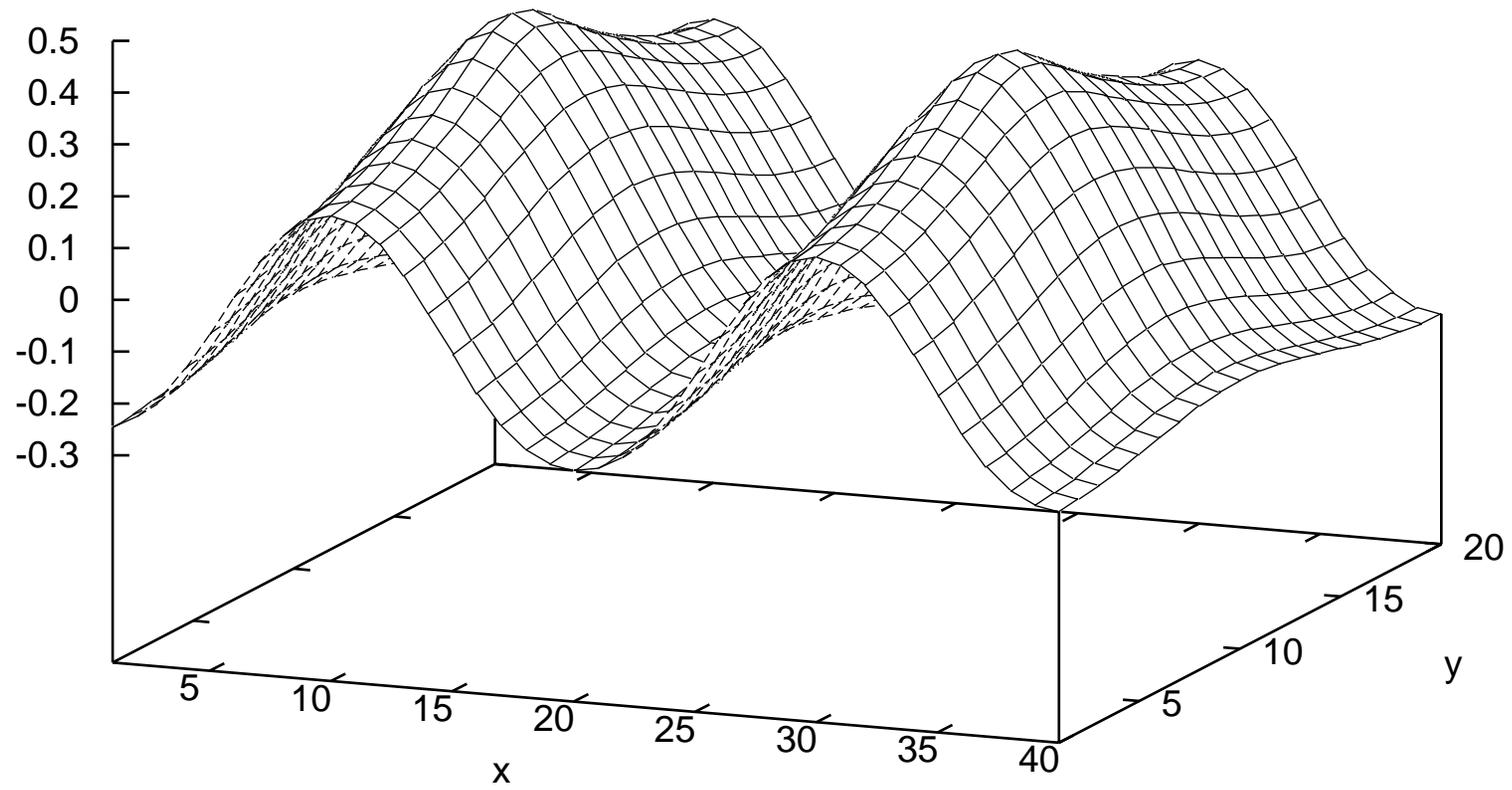
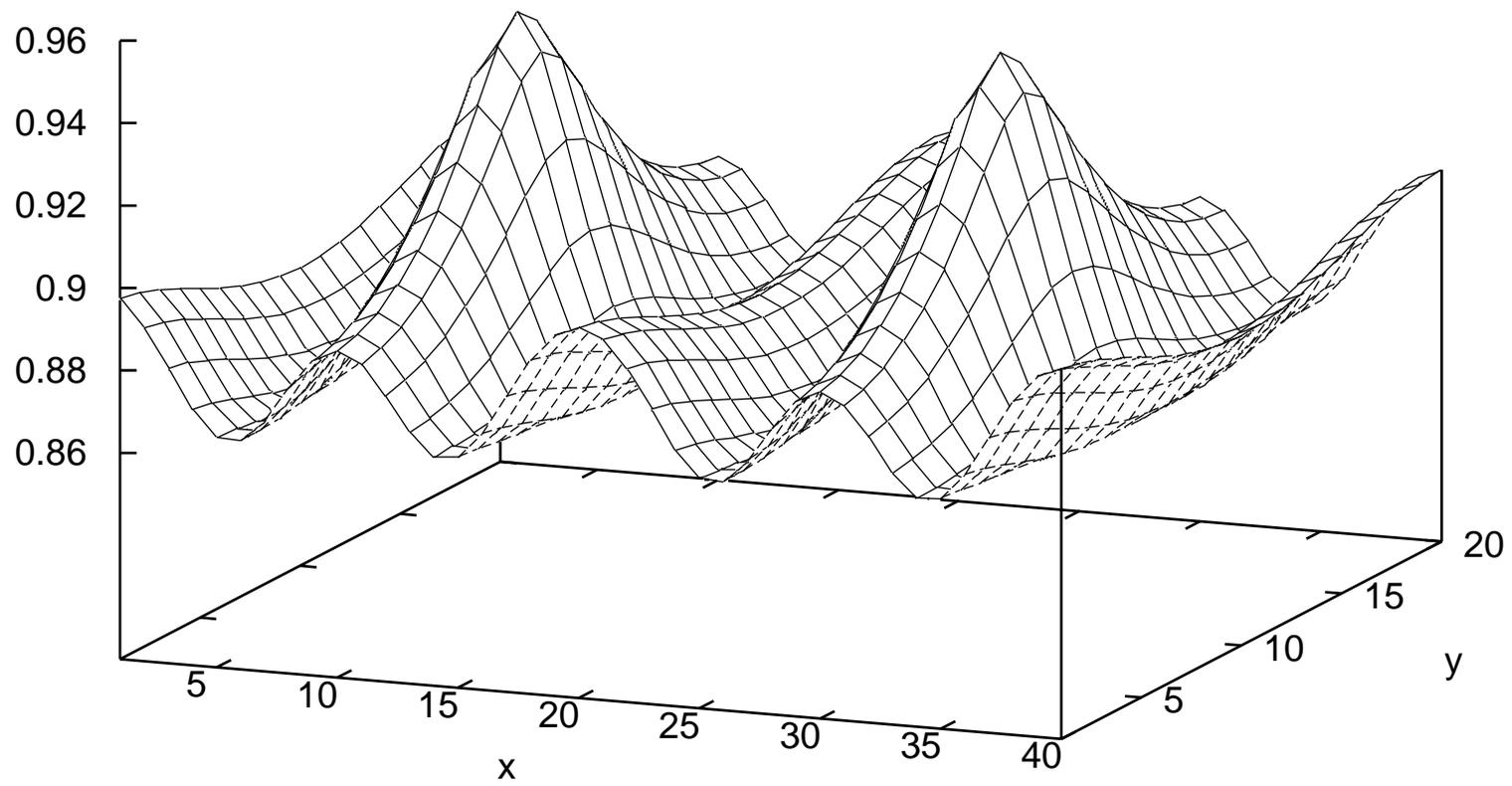
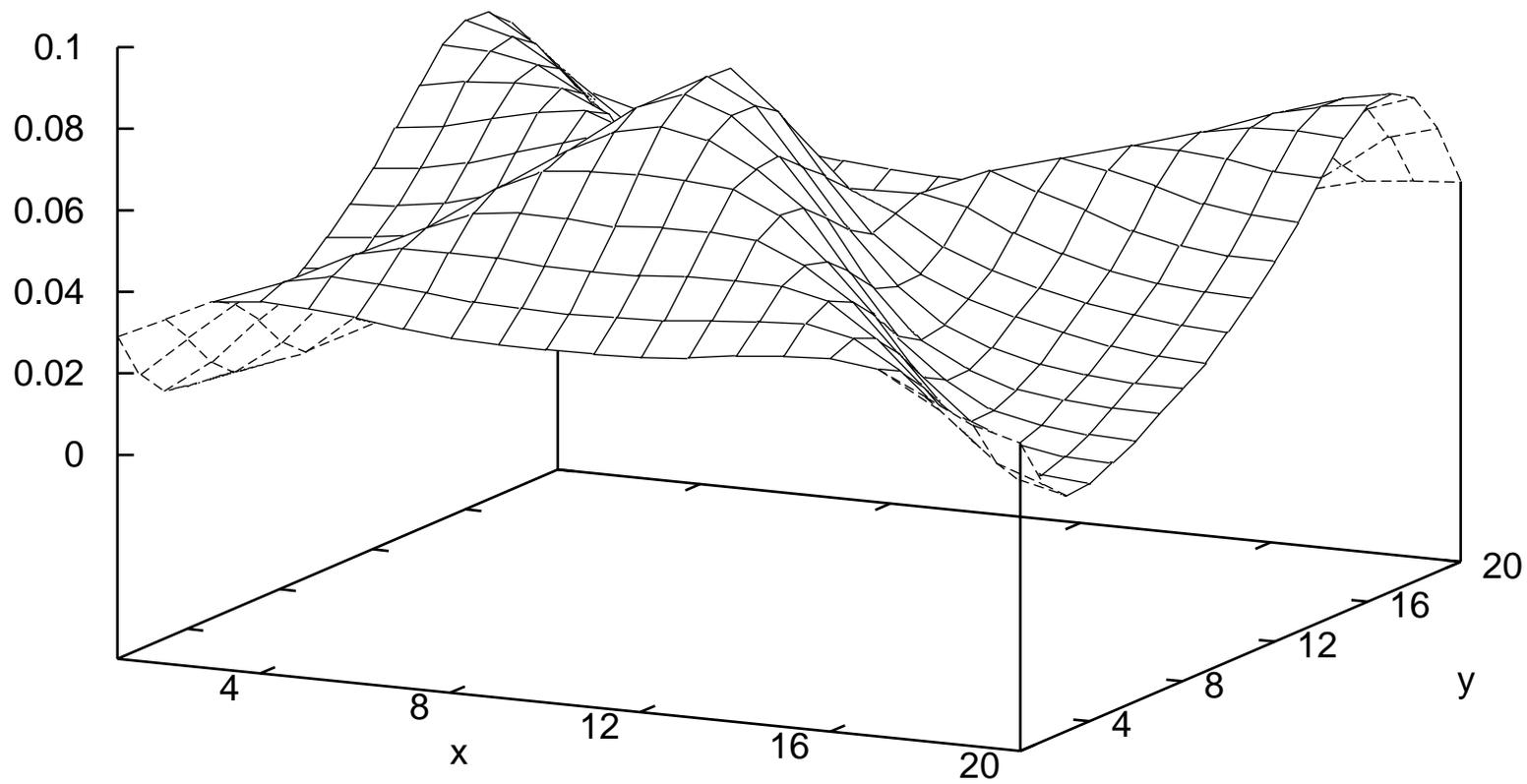
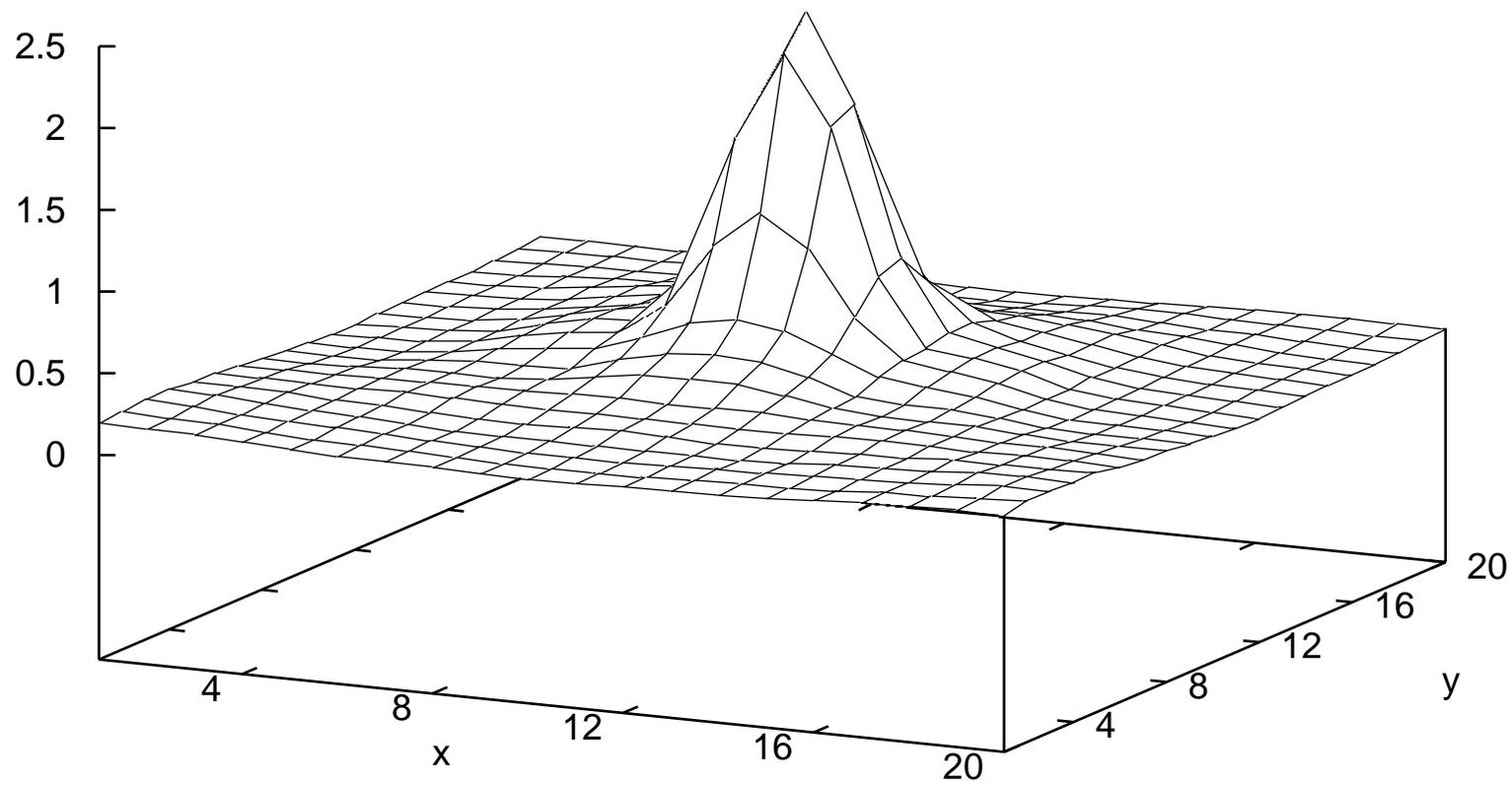
\end{document}